\documentclass[twocolumn,epjc3]{svjour3}
\journalname{Eur. Phys. J. C}
\RequirePackage{graphicx}
\usepackage[utf8]{inputenc}
\usepackage{hyperref}
\usepackage{subcaption}
\usepackage{upgreek}
\usepackage{siunitx}
\usepackage{booktabs}
\usepackage{color} 
\usepackage[normalem]{ulem}
\graphicspath{{figures/}}

\begin{document}

\title{An active transverse energy filter to differentiate low energy particles with large pitch angles in a strong magnetic field}

\titlerunning{\parbox{9cm}{An active transverse energy filter}} 
\authorrunning{K.~Gauda, \dots, C.~Weinheimer}
%authors
\institute{%
Institute for Nuclear Physics, Westf\"{a}lische Wilhelms-Universit\"{a}t M\"{u}nster, Wilhelm-Klemm-Str. 9, 48149 M\"{u}nster, Germany\label{a}
\and Institute for Astroparticle Physics~(IAP), Karlsruhe Institute of Technology~(KIT), Hermann-von-Helmholtz-Platz 1, 76344 Eggenstein-Leopoldshafen, Germany\label{b}
\and Institute of Experimental Particle Physics~(ETP), Karlsruhe Institute of Technology~(KIT), Wolfgang-Gaede-Str. 1, 76131 Karlsruhe, Germany\label{c}
\and CeNTech and Physics Institute, Heisenbergstr. 11, Westf\"{a}lische Wilhelms-Universit\"{a}t M\"{u}nster, Germany\label{e}
\and Kirchhoff-Institute for Physics, University of Heidelberg, Im Neuenheimer Feld 227, 69120 Heidelberg, Germany\label{f}
\and Center for Experimental Nuclear Physics and Astrophysics, and Dept.~of Physics, University of Washington, Seattle, WA 98195, USA\label{g}
}
\thankstext{email1}{e-mail: k.gauda@uni-muenster.de}
\thankstext{email2}{e-mail: hannen@uni-muenster.de}
\thankstext{email3}{e-mail: weinheimer@uni-muenster.de}
% Authors:
\author{%
K.~Gauda\thanksref{a,email1}
\and S.~Schneidewind\thanksref{a}
\and G.~Drexlin\thanksref{b,c}
\and A.~Fulst\thanksref{a}
\and V.~Hannen\thanksref{a,email2}
\and T.~König\thanksref{a}
\and A.~Lokhov\thanksref{a}
\and P.~Oelpmann\thanksref{a}
\and H.-W.~Ortjohann\thanksref{a}
\and W.~Pernice\thanksref{e,f}
\and R.G.H.~Robertson\thanksref{g}
\and R.W.J.~Salomon\thanksref{a}
\and M.~Stappers\thanksref{e}
\and C.~Weinheimer\thanksref{a,email3}
}

\date{\today}
\maketitle

\begin{abstract}
We present the idea and proof of principle measurements for an angular-selective active filter for charged particles. The motivation for the setup arises from the need to distinguish background electrons from signal electrons in a spectrometer of MAC-E filter type.
While a large fraction of the background electrons exhibit predominantly small angles relative to the magnetic guiding field (corresponding to a low amount of kinetic energy in the motion component transverse to the field lines, in the following referred to as transverse energy) and pass the filter mostly unhindered, signal electrons from an isotropically emitting source interact with the active filter and are detected. The concept is demonstrated using a microchannel plate (MCP) as an active filter element. When correctly aligned with the magnetic field, electrons with a small transverse energy pass the channels of the MCP without interaction while electrons with large transverse energies hit the channel walls and trigger an avalanche of secondary electrons that is subsequently detected.  
Due to several drawbacks of MCPs for an actual transverse energy filter, an alternative detection technique using microstructured Si-PIN diodes is proposed. 
\end{abstract}
\section{Introduction}
\label{sec:intro}
\begin{figure}[h!]
 \centering 
 \includegraphics[width=\columnwidth]{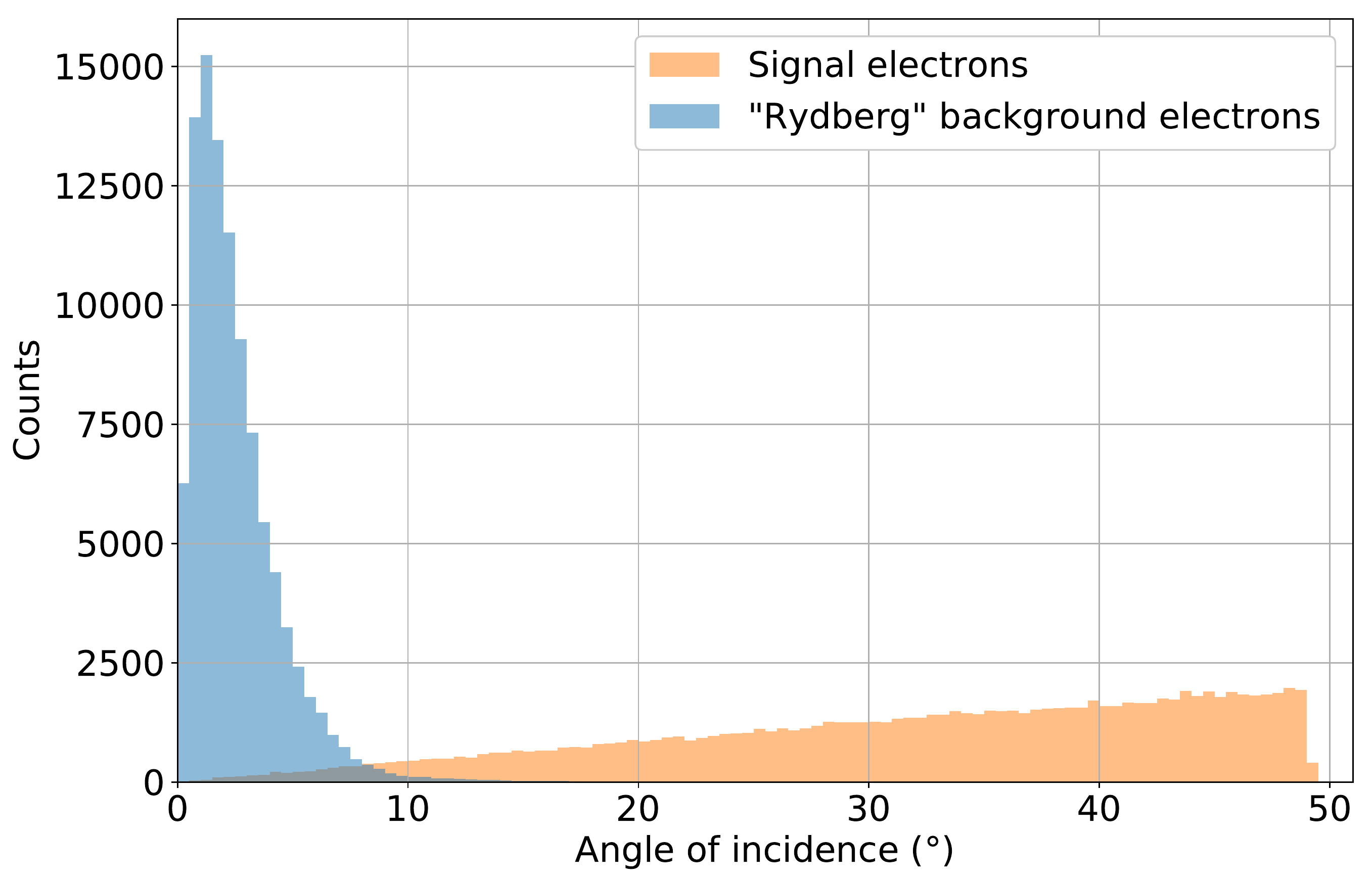}
 \caption{Angular distributions of signal (orange) and ``Rydberg'' background electrons (blue) at the detector of the KATRIN experiment simulated for $10^5$ electrons using input from~\cite{ref:Trost}. The simulation considers the energy spectrum of background electrons which are ionized by black-body radiation at room temperature from hydrogen (36\,\%) and oxygen atoms (64\,\%) in highly excited Rydberg states. These Rydberg atoms are sputtered from the spectrometer walls % with a $\cos(\theta)$-distribution 
 due to $\alpha$-decays of implanted $^{210}$Pb progenies from the $^{222}$Rn decay chain~\cite{ref:Rydberg}.
 The Rydberg atom's energies from \cite{ref:Trost} are corrected for the surface binding energy and -- after ionisation -- the electron energy and direction are calculated with the Doppler shift of the Rydberg atoms taken into account.}
 %Energien der Atome an der Grenzfläche Festkörper-Vakuum aus der Doktorarbeit um die Austrittsarbeit korrigieren und damit dann nach der Ionisation Energie und Richtung der Elektronen aufgrund der Dopplerverschiebung berechnen}
 \label{fig:background_at_KATRIN} 
\end{figure}

In rare-event searches or experiments where a tiny signal is to be measured, an efficient suppression of background events by a suitable detector is essential. Often, signal events in detectors can be distinguished from background events by different energy deposits, by different specific energy losses or other characteristic quantities \cite{Wermes}. However, there are cases in which signal and background events are caused by the same type of particle, arriving with a very similar energy and differing only by their angle of incidence. Standard methods to distinguish different angles of incidence are low-density tracking detectors (e.g. gas-filled drift chambers) or a $\Delta E$-$E$ arrangement\footnote{In the latter, the energy loss $\Delta E$ of an ionizing particle in a thin detector in front of the main detector, in which the particle deposits its remaining energy $E$, is a measure of the angle of incidence.}. These methods are not applicable if the energy of the particle is too low, or if other requirements like ultra-high-vacuum compatibility do not allow the use of gas detectors with ultra-thin entrance windows.
Another contraindication is given by tracking challenges if the particle path is bent by a strong magnetic field into a spiral path with a cyclotron radius well below a millimeter. \\
All these restrictions apply to the Karlsruhe Tritium Neutrino experiment KATRIN~\cite{ref:TDR2}, which aims to determine the neutrino mass from a very precise measurement of the electron energy spectrum of molecular tritium beta decay near its endpoint of 18.57\,keV. A large (10\,m diameter, 23\,m length) spectrometer of MAC-E-filter~\cite{mac-e} type allows electrons above an adjustable threshold with $\cal{O}$(1\,eV) width to be transmitted to an electron detector.
This detector -- a rear-illuminated Si-PIN diode of 9\,cm diameter segmented into 148 pixels of equal area~\cite{FPD} -- is placed inside a superconducting solenoid within a magnetic field of 2.4\,T.
The signal electrons gyrate around the magnetic field lines on their way from the spectrometer to the detector and arrive at the detector with a maximal angle of incidence of about 50$^\circ$ to the surface normal (not taking into account post acceleration). This corresponds to a cyclotron radius at the detector of less than $200\,\upmu$m\footnote{In our brief description we omit that the electrons are accelerated in front of the detector by 10\,keV using a post-acceleration electrode modifying the angular distributions in fig.~\ref{fig:background_at_KATRIN}. This fact can be neglected here because it only adds longitudinal energy without changing the cyclotron radius.}.\\
In contrast to the signal electrons, a significant share of the background electrons in the KATRIN spectrometer originate from highly excited, so-called Rydberg atoms, which are ionized by black-body radiation emitted from the spectrometer walls at room temperature~\cite{ref:Rydberg}. These secondary electrons possess very low starting energies~\cite{ref:Trost}, but are accelerated by the electric potential gradient of the spectrometer and arrive at the detector with essentially the same kinetic energy as the signal electrons (which in the central region of the spectrometer also possess energies in the eV range). However, since the initial transverse energy of the secondary emitted electrons is only ${\cal O}(k_B T \approx 25\,\mathrm{meV})$ and they do not acquire transverse energy in the electric field gradient towards the exit of the spectrometer due to the strong magnetic field there, they reach the detector with incidence angles of typically less than 10° to the surface normal, see fig. \ref{fig:background_at_KATRIN}. 
After initial measures to reduce this background component have been successfully implemented~\cite{ref:SAP}, KATRIN's background is still an order of magnitude higher than anticipated. Therefore, we propose a method which should enable further background reduction by utilizing the different angular distributions of signal and background electrons at the detector.

The following section provides an introduction to the idea of an angular selective detector and a possible realization of the device using a microchannel plate. Section~\ref{sec:test} discusses a proof of principle measurement of the idea at a test setup in Münster, while section~\ref{sec:implementation} presents Monte-Carlo simulations for an aTEF design adapted to the needs of the KATRIN experiment.
\section{The aTEF idea}
\label{sec:idea}
\begin{figure}[t]
 \centering 
 \includegraphics[width=\columnwidth]{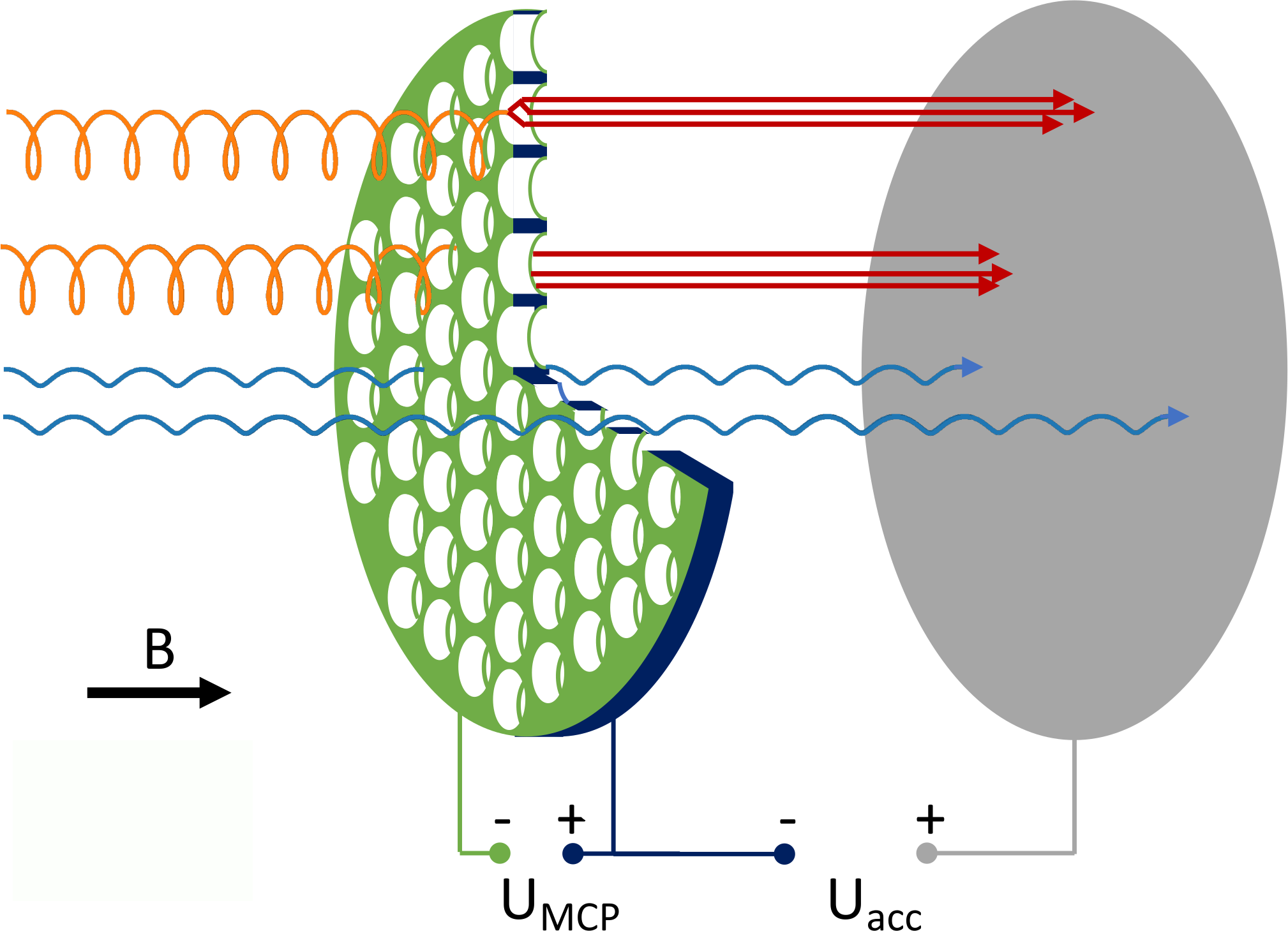}
 \caption{Principle of an MCP-based active transverse-energy filter in front of an electron detector (gray) in a horizontal magnetic field $B$. Electrons with a small cyclotron radius (blue) pass the MCP  without interaction in most cases, hit the detector and get counted there. These electrons giving rise to a standard energy deposition in the detector are considered to be mostly background electrons (see Fig.~\ref{fig:background_at_KATRIN}).
 Electrons with a large cyclotron radius (orange) have a large chance to hit the MCP walls since they possess significant pitch angles of typically $\ge 10^\circ$ and possess incidence angles with respect to the MCP channel normal of typically $\le 80^\circ$. Thus, they possess a large chance to give rise to a cascade of secondary electrons (red) that are accelerated via $U_\mathrm{acc}$ to the detector and lead to a large signal there. In order to tag the primary electrons with large cyclotron radius -- the goal of aTEF, those electrons producing large detector signals will be classified as signal electrons (see Fig.~\ref{fig:background_at_KATRIN}). }
 \label{fig:mcp_atef_principle} 
\end{figure}

We propose here -- in the special situation of low-energy charged particles in strong magnetic fields, such as the electrons in the KATRIN experiment -- to geometrically distinguish the impact angles or the transverse energy of the particles via the cyclotron radius of the spiralling motion around the magnetic field lines. \\
A suitable device for this purpose could be a microchannel plate (MCP). Typically, when charged particles such as electrons or ions 
%or even neutral particles such as atoms or photons 
hit the side walls of a channel with diameter $d$ within the MCP  they trigger the emission of secondary electrons from the wall, which is coated with a material with a particularly high secondary-electron yield. The number of electrons ejected in these interactions also depends on type, energy and angle of the incoming particle. The secondary electrons are accelerated by the electric field applied across the channel, presumably hit the wall again and, thus, trigger a secondary electron cascade \cite{ref:MCP}.\\
In commercial MCPs the channels are tilted by 8°-15°~\cite{ref:MCP} with respect to the surface normal. 
In figure~\ref{fig:mcp_atef_principle} we purposely assume channels parallel to the magnetic field lines, since untilted channels allow electrons with small cyclotron radii to pass the MCP without interaction. Thereby, a distinction between electrons depending on their angle to the channel axis becomes possible.
%\sout{We deliberately neglect this tilt angle of the channels in fig. \ref{fig:mcp_atef_principle}.} 
%
If a charged particle with a small cyclotron radius $r_\mathrm{c} \ll d/2$ (blue spiral track) enters a channel of the MCP, it usually passes through unaffected and causes a small signal in the detector (here in gray). If, on the other hand, a charged particle with a large cyclotron radius $r_\mathrm{c} > d/2$ (orange spiral track) enters a channel, the particle will most likely hit the channel walls and trigger a secondary electron avalanche (red arrows). An applied voltage $U_\mathrm{acc}$ will accelerate the electron avalanche towards the detector where it will cause a comparably larger signal. If the analysis discriminates on the measured signal height, the channels thus act as a ``Transverse Energy Filter''. Given that the filter is not a passive element but aids in the detection of the signal electrons, we name this principle ``active-Transverse Energy Filter'' (aTEF).\\
We have explained the aTEF idea using the example of an MCP. In general, this principle can be realized on any kind of geometric aperture that is instrumented with a suitable method of particle detection. We will see in section~\ref{sec:implementation} that a suitable aperture geometry can be a honeycomb structure due to its ideal open area ratio. The instrumentation of such a honeycomb structure could be realized not only with a secondary-electron-emitting layer but also by using scintillator or semiconductor materials to generate photons or electron-hole pairs and thereby detectable signals.
\section{Proof of the aTEF principle with a microchannel plate}
\label{sec:test}
\begin{figure*}[h]
 \centering 
 \includegraphics[width=\textwidth]{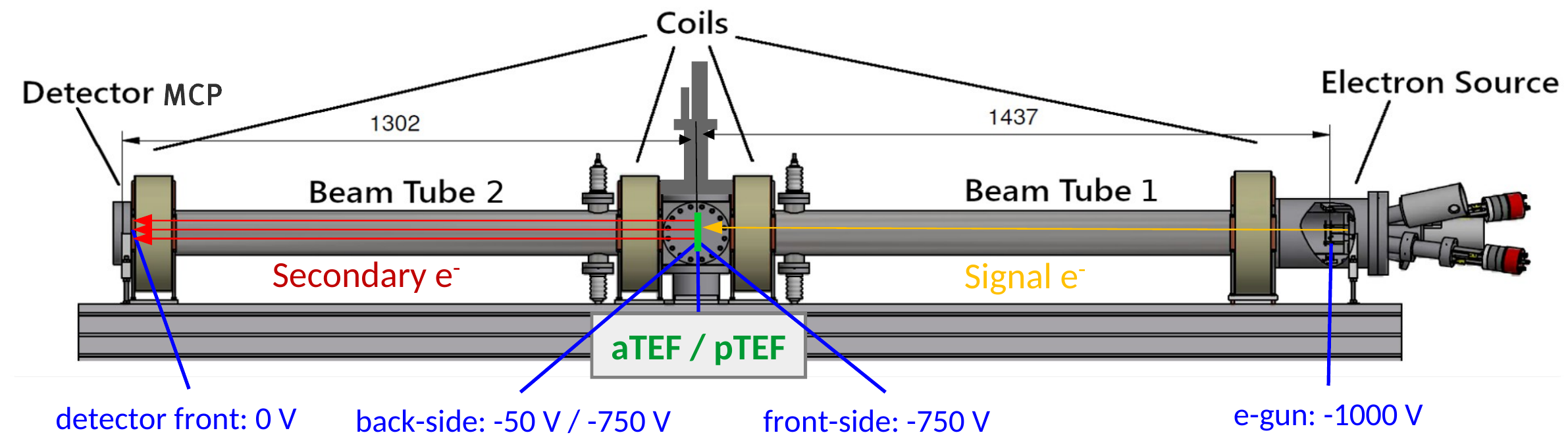}\\[1mm]
 \includegraphics[width=\textwidth]{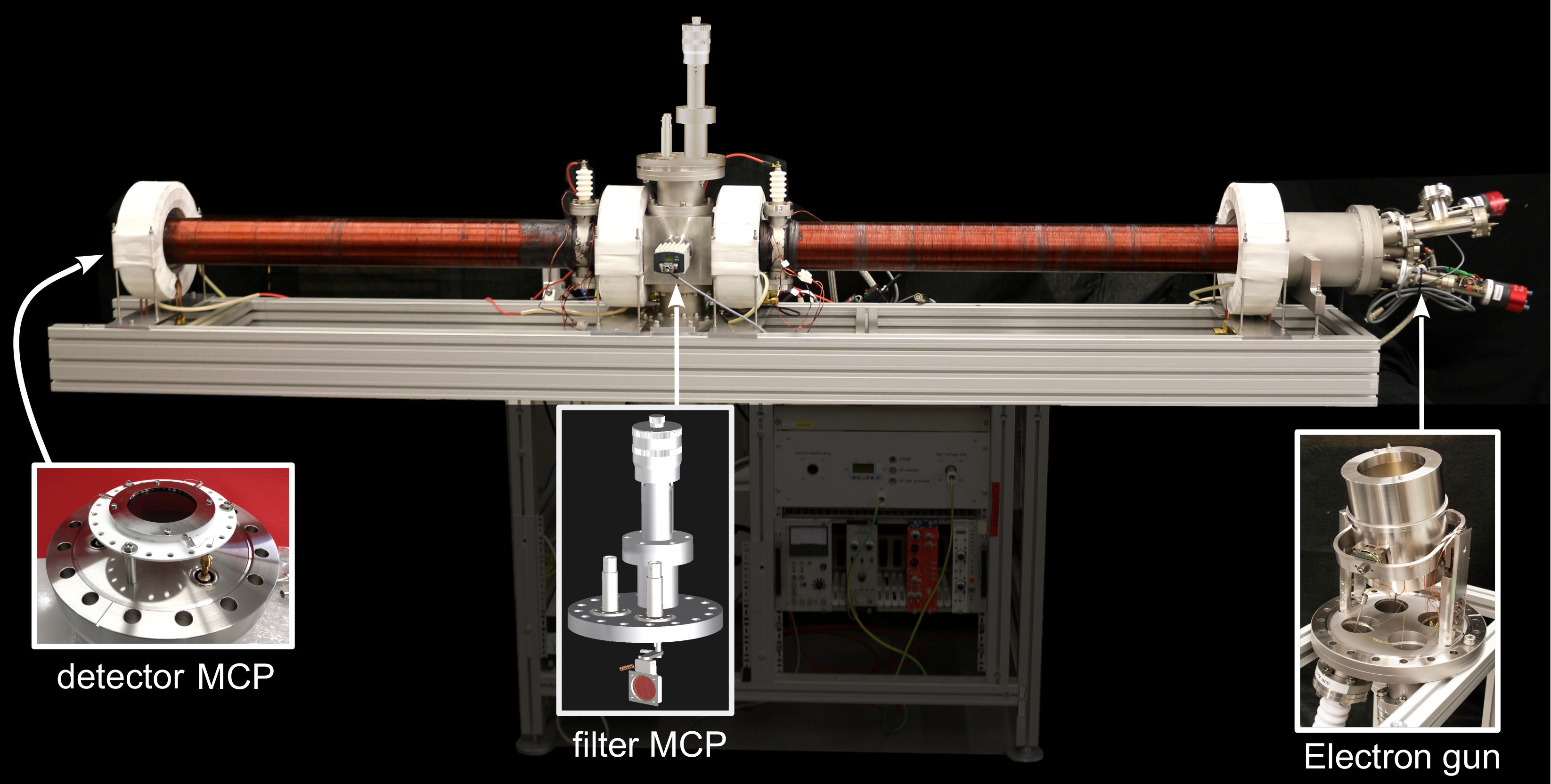}
 \caption{Upper panel: Schematic view of the test setup with  detector MCP, solenoid coils, beam tubes, central vacuum chamber with transverse-energy filter and electron source. Lower panel: Photo of the actual setup.}
 \label{fig:tof_setup} 
\end{figure*}

\begin{figure*}[h]
 \centering 
 \includegraphics[width=\textwidth]{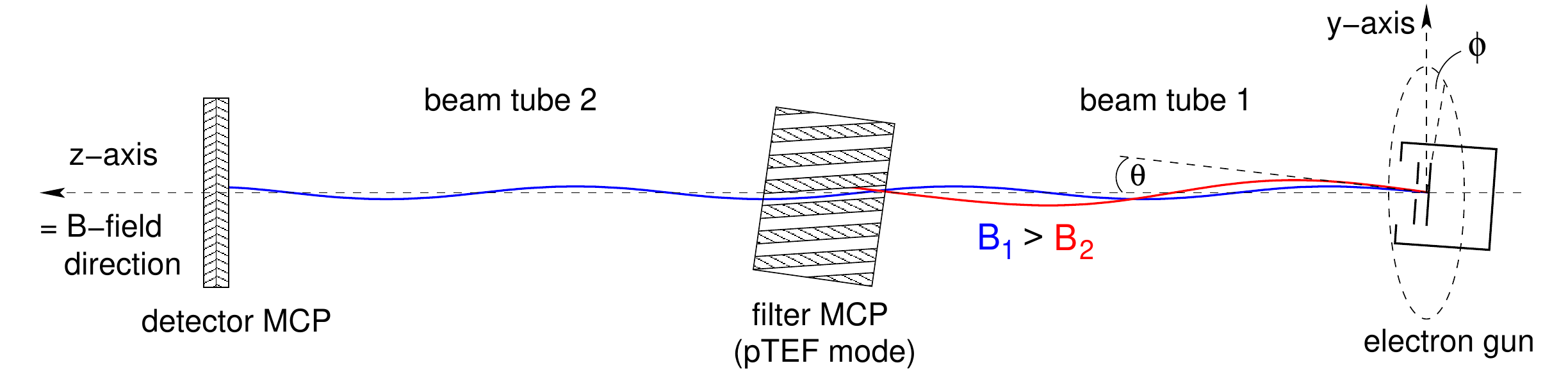}
 \caption{Electron tracks in pTEF mode at two different magnetic-field strengths in beam tube 1 (drawing not to scale). Due to residual misalignments of the channels of the filter MCP and the magnetic field direction (parallel to the z-axis), a suitable combination of electron-gun angles (polar angle $\theta$ and azimuthal angle $\phi$) and magnetic field strength $B$ in beam tube 1 is required for the electrons to pass the filter ($B_1$, blue track). Changing the magnetic field changes the phase and thus the angle with which the electrons enter the channels of the filter MCP and causes them to be absorbed in the pTEF mode ($B_2$, red track).}
 \label{fig:angles} 
\end{figure*}

\begin{figure}[h]
 \centering 
 \includegraphics[width=\columnwidth]{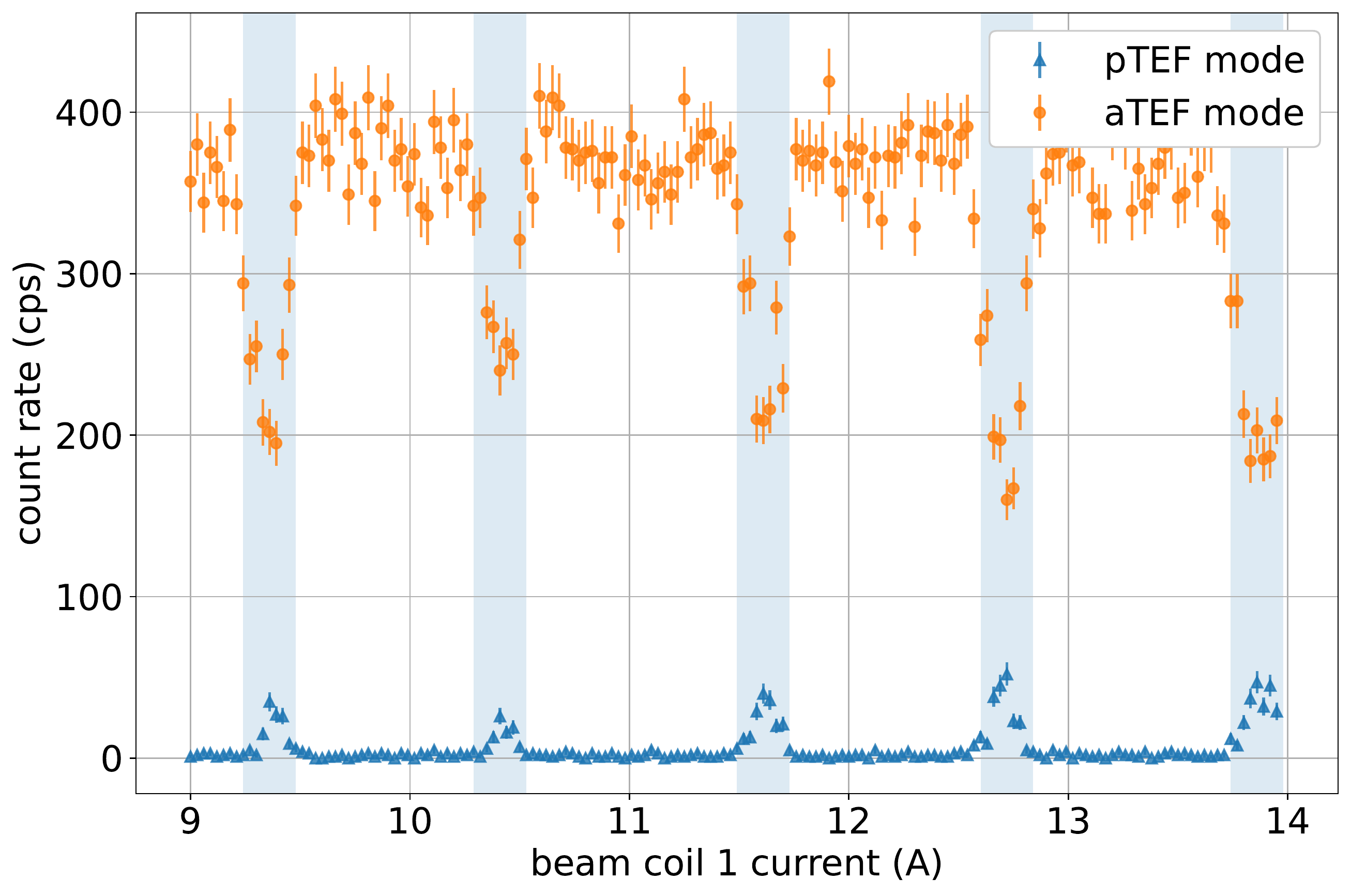}
 \caption{Detected count rates as functions of beam coil~1 current when operating the transverse-energy filter in active mode (orange dots) or passive mode (blue triangles).}
 \label{fig:rates} 
\end{figure}

As a proof of principle of the ideas described in section~\ref{sec:idea}, measurements have been performed using a microchannel plate (MCP) by RoentDek\footnote{RoentDek Handels GmbH, Im Vogelshaag 8, 65779 Kelkheim, Germany} with an active diameter of 25\,mm and an open-area ratio (OAR) of 60\,\% as an active (aTEF) or passive (pTEF) filter element. The circular channels inside the plate have a diameter of $10\,\upmu$m, a channel depth of $400\,\upmu$m and a bias angle of $12^\circ$. This filter MCP was inserted into the central vacuum chamber of a test setup constructed originally for Time-Of-Flight (TOF) measurements with electrons from a pulsed, angular-selective photoelectron source~\cite{ref:egun}, see fig.~\ref{fig:tof_setup}.
Electrons with kinetic energies of 1\,keV were created in the source and were guided by six solenoid coils through the filter and towards a detector at the opposite side of the setup. The setup contains two types of magnets: Four cooled {\em anoxal}\,\textsuperscript{\tiny\textregistered}\footnote{Umwelttechnik Wesselmann GmbH, Auf dem Knuf 21, 59073 Hamm, Germany} (anodized oxidized aluminum) coils operated at about \SI{40}{\milli\tesla} and two long solenoid coils wound around the beam tubes 1 and 2 and operated at about \SI{7}{\milli\tesla}. The electron detector consists of an MCP stack in chevron configuration with an active diameter of \SI{46}{\milli\meter} and an open-area ratio of 60\% (type MCP-050-D-L-A-F from tectra GmbH\footnote{tectra GmbH, Reuterweg 51 – 53, 60323 Frankfurt/M, Germany}) mounted on a CF100 flange and operated at a back plate voltage of 1853\,V. Signals from the MCP stack were amplified using an Ortec\footnote{ORTEC GmbH, Am Winterhafen 3, 28217 Bremen, Germany} 474 timing filter amplifer set to a gain of 20 and a shaping time of either \SI{20}{\nano\second} for detection of single electrons (passive mode) or 500\,ns for the detection of multiple electrons created in the active-filter mode. The amplified signals were then discriminated by a CAEN\footnote{CAEN S.p.A., Via Vetraia, 11, 55049 – Viareggio (LU), Italy} N417 leading-edge discriminator set to a threshold of \SI{100}{\milli\volt} before counting.\\
The filter MCP was operated in two different modes: either as an active transverse energy filter with a potential difference of \SI{700}{\volt} applied between front (source side) and back (detector side), or as a passive filter element with equal voltages applied to both sides (see settings in table~\ref{tab:voltages}).
\begin{table*}[h]
    \centering
        \caption{Voltages applied to the filter MCP and shaping time used for the detector MCP signals in the active (aTEF) and passive (pTEF) filter modes.}
    \begin{tabular}{crrr}
    \toprule
                  & front-side voltage (V) & back-side voltage (V) & shaping time (ns)\\ \midrule
        aTEF mode & -750             & -50              & 500\\
        pTEF mode & -750              & -750          & 20\\
        \bottomrule
    \end{tabular}
    \label{tab:voltages}
\end{table*}
The front-side voltage is identical in both cases in order to decelerate the electrons in the same manner before entering the filter MCP in both the active (aTEF) and the passive (pTEF) mode.
The filter MCP was mounted on a retractable feedthrough so that its channels were aligned with the z-axis (identical with the magnetic field axis) of the setup. \\
Due to residual misalignments between the channels and the magnetic-field lines in the central mounting position, a suitable orientation of the photoelectron source that allowed for unhindered transmission of electrons through the channels had to be determined in the passive filter mode prior to the measurements (see figure~\ref{fig:angles}). 
For this purpose the timing filter amplifer was set up for single-electron detection and the polar and azimuthal angles of the electron source were scanned around the zero position to maximize the number of electrons transmitted through the filter. The maximum count rate was observed at a polar angle of $4^\circ$ and an azimuthal angle of $150^\circ$ of the electron gun. 
Due to the non-zero polar angle at this setting, the electrons perform a cyclotron rotation around the magnetic field lines with the initial phase determined by the azimuthal angle. 
At the filter MCP in the center of the setup we have a magnetic field strength of about 13.8\,mT, a cyclotron frequency of $f_c = e\, B /(2\,\pi\,m_e) = 386\,$MHz and a corresponding cyclotron radius of $\approx 190\,\upmu$m at an electron polar angle of $3^\circ$. While this radius is much larger than the channel diameter of $10\,\upmu$m, the transversal motion during the passage of the electrons inside the channels amounts only to $20\,\upmu$m. At a channel length of $400\,\upmu$m this can be compensated if the electrons enter the channel under an angle of $\approx 2.9^\circ$ with the correct cyclotron phase.
As shown in figure~\ref{fig:angles}, we can vary the cyclotron frequency in beam tube 1 and thus the cyclotron phase and the angle with which the electrons enter the filter MCP by changing the field strength in the coil wound around beam tube 1. Varying the coil's current therefore causes a modulation of the number of electrons which can pass the channels unhindered as evident in the lower dataset (blue triangles) of fig.~\ref{fig:rates}. 
%
%Here the count rate at the detector is given as function of the current in beam tube coil~1. 
The observed maxima correspond to electrons hitting the filter MCP with the correct angle to pass through the channels. The spacing of the maxima is $\Delta I_{\rm exp} = (1.13 \pm 0.05)\,$A and corresponds to the current change required for a full rotation of the phase angle. This value is in good agreement with the predicted value $\Delta I_{\rm theo} = 1.12\,$A, deduced by approximating the magnetic field inside the evacuated beam tube by $B \approx \mu_0 \cdot I \cdot n$, with $\mu_0$ being the vacuum permeability and $n=0.457\,{\rm mm}^{-1}$ the winding density of the coil. \\
When operating the filter MCP in active transverse-energy-filter mode, we are interested in the detection of secondary electron cascades caused by primary electrons hitting the MCP walls. Thus, we have increased the energy threshold of the detector substantially and enlarged the shaping time of the timing filter amplifer to 500\,ns.
In this mode the dataset given by the orange dots in fig.~\ref{fig:rates} is obtained. Here the dips in count rate correspond to electrons traversing the MCP channels without or with only weak interaction yielding no or a tiny secondary electron cascade only. The plateaus are created by electrons that hit the channel walls under non-grazing incidence and create  avalanches of secondary electrons which are then detected. 
The mean rate on the plateaus is $(370 \pm 21)\,$cps and, assuming that the primary electrons that interact with the channel walls create a secondary electron signal that is always detected~\footnote{The open-area ratio of the detector MCP does not reduce the number of avalanche events that are detected, as we can safely assume that not all secondary electrons of an avalanche will be blocked by the material between the channels.}, this rate corresponds to the \SI{60}{\percent} fraction of electrons created in the photoelectron source which enter the open area of the filter MCP.\\
With a suppression efficiency of about \SI{80}{\percent} for the detection of single electrons (due to the longer shaping time in the aTEF mode) we would expect the rate dips observed in the aTEF mode to go down to a value of $0.2 \cdot 370\,{\rm cps} = 74\,$cps, while we actually observe a drop to a rate of $\approx 200\,$cps only. 
Likewise we would expect peak rates up to $0.6 \cdot 370\,{\rm cps} = 222\,$cps in passive mode (here we have to take into account the 60\% open area ratio of the detecting MCP) when all primary electrons that enter a channel in the filter MCP pass unhindered, but observe only a maximum rate of $\approx 50$\,cps. 
%
%There are two main reasons we do not observe the expected event rates. First, even an electron with a cyclotron radius smaller than the channel radius has a significant probability to hit the channel wall because we do not have just an aperture but a long channel (see also Fig.~\ref{fig:aTEF_simulation}). In our case the channels have an unfavorable diameter-to-length ratio of 10:400. Secondly, the angular distribution of electrons produced in the photoelectron source has a non-zero width of $\sigma \approx 0.8^\circ$ at the channel plate and the path of these electrons through the channel has a curvature determined by their polar angle, the magnetic-field strength and their velocity at the position of the filter MCP. 
The main reason we do not observe the expected event rates is the unfavorable diameter-to-length ratio of the channels of 10:400, which means that only electrons entering with precisely the right angle are able to pass the filter unhindered. Given that the angular distribution of the electrons created by the electron gun has a non-zero width of $\sigma \approx 0.8^\circ$ at the filter MCP, many of these electrons will hit the channel walls even at the optimum angular settings. These events are then lost for detection in the passive mode and contribute to the observed rate deficit, while causing an increased rate in the dips of the active mode spectrum where these electrons create avalanches that are detected.\\
To construct an efficient transverse-energy filter we therefore have to optimize the geometry of the device, most notably the diameter-to-length ratio and the open-area ratio.

\section{Geometry of an aTEF with high efficiency for the KATRIN experiment}
\label{sec:implementation}
\begin{figure*}[t]
  \centering
  \includegraphics[width=0.64\textwidth]{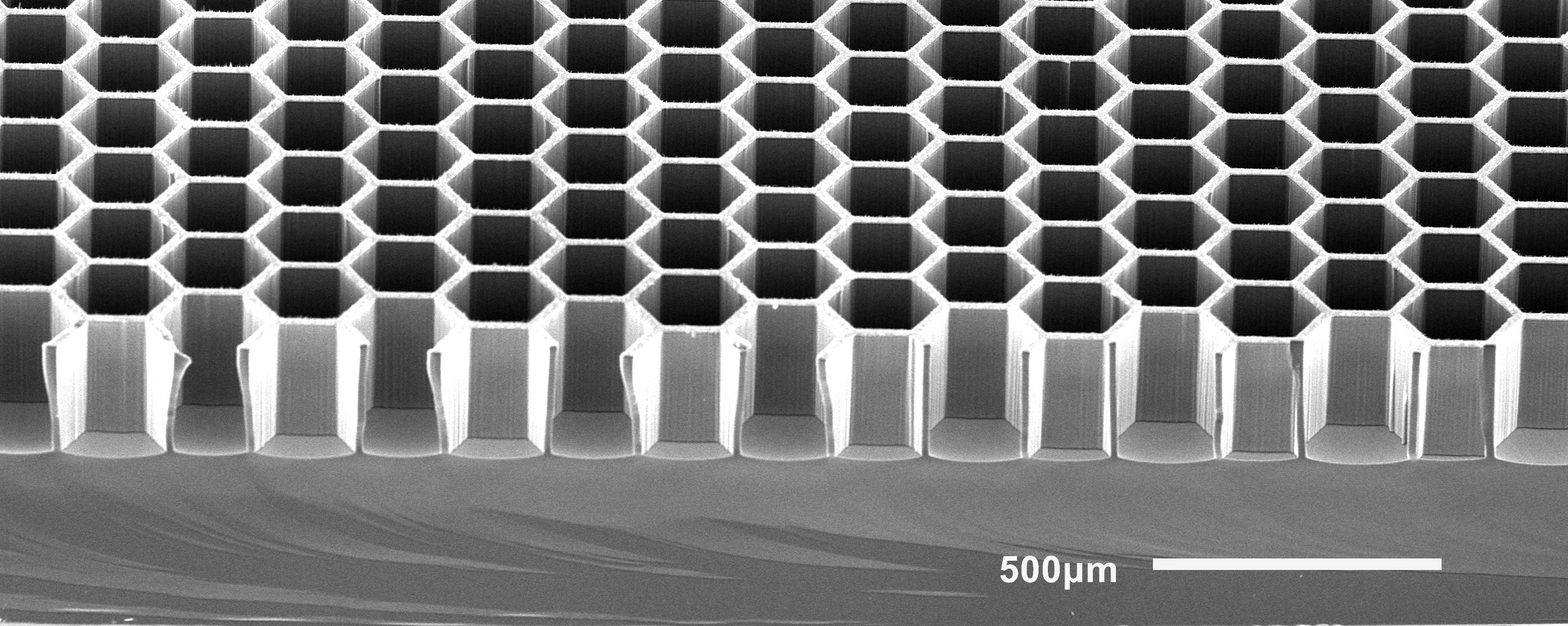}\hfill
  \includegraphics[width=0.34\textwidth]{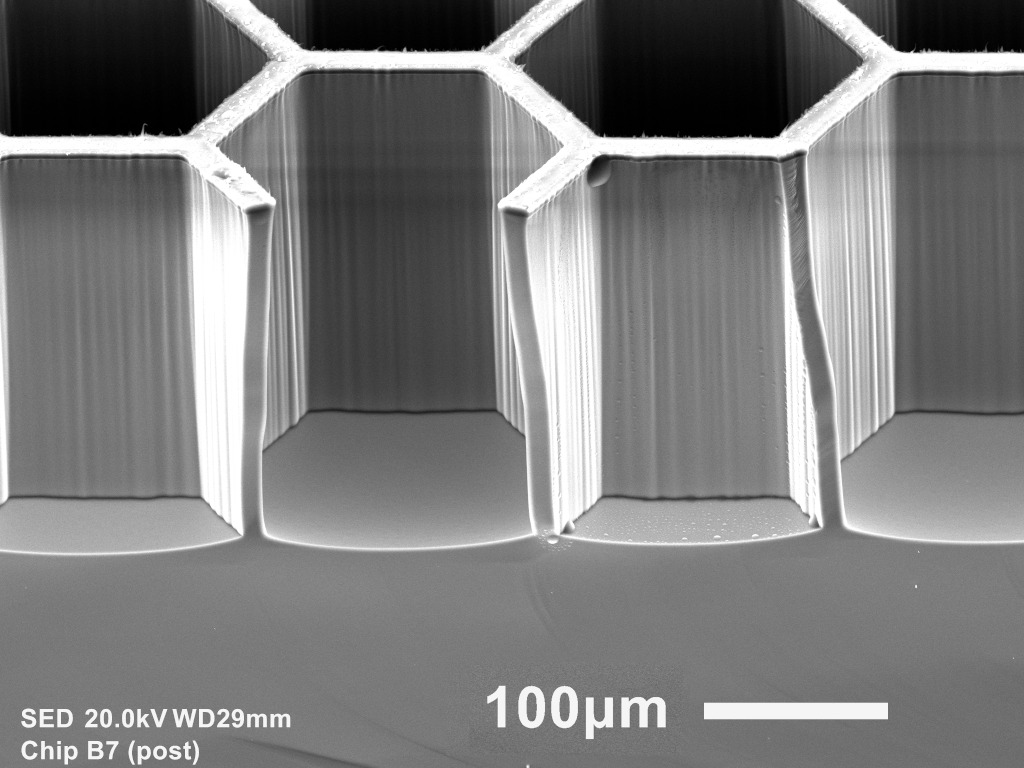}
  \caption[caption]{Images of a silicon microstructure fabricated in the Münster Nanofabrication facility\footnotemark~via ICP-RIE, taken with a JEOL\footnotemark~JSM-IT100 scanning electron microscope. The hexagon side length is $100\,\upmu$m, wall thickness $10\,\upmu$m and etched depth $225\,\upmu$m.}
\label{fig:etchresult}
\end{figure*}

Commercial MCPs like the one used for a proof of principle test in the setup come along with properties that rule out their use as an ``aTEF'' in a low-rate counting experiment with strong magnetic fields like KATRIN, including typical channel radii, open-area ratios, aspect ratios and dark count rates.
A self-manufactured or manipulated detector -- not necessarily of MCP type -- may overcome these drawbacks. \\
For a suitable design, we have to adapt the diameter-to-length ratio of the channels to allow background electrons with small transverse energy to pass the filter unhindered while most of the signal electrons are detected. A channel bias angle is disfavored here (unlike in typical commercial MCPs where it is mainly used to reduce ion feedback~\cite{MCP-angle}), since the channels for the proposed ``aTEF'' have to be parallel to the magnetic-field lines.
Furthermore, the open-area ratio needs to be maximized to reduce losses due to signal electrons being blocked by the material in between the channels.
Using a hexagonal channel shape instead of the circular shape of regular MCPs allows an open-area ratio of up to $90\,\%$ (open-area ratios of \SI{83}{\percent} with hexagonal channels have already been reported~\cite{MCP-ALD}).\\
To overcome the inherent dark-count rate of the glass materials usually used for MCP production, silicon as an inherently radiopure material with an intrinsic background rate $<\SI{0.01}{cps\per cm^2}$ can alternatively be used to manufacture MCPs~\cite{Si-MCP2,Si-MCP} but also allows for completely different detection mechanisms as e.g. used in Si-PIN diodes\footnote{If a Si-PIN diode would still drift electron-hole pairs after deep cryo-etching, then such a hexagonal structure could be etched directly into the electron detector (grey disk in Fig. 2) of the KATRIN experiment, which is a backside-illuminated pixelized Si-PIN diode. The (signal) electrons with large cyclotron radii would hit the side walls of the hexagonal structure and create electron-hole pairs there, which should then be drifted to the electrodes. An 18.6 keV electron at the endpoint of the tritium beta spectrum at KATRIN would create about 5000 electron-pairs, enough to produce a measurable charge signal if the etching process would only moderately increase the detector leakage or other noise sources. The (background) electrons with small cyclotron radii would hit the bottom of the structure, which would need to be passivated, and would not generate a signal.}\\
The previous considerations lead to a filter design with hexagonal channels with a side length of $100\,\upmu$m, a channel-wall thickness of $10\,\upmu$m and a depth of $400\,\upmu$m. The targeted geometry can be manufactured in silicon via a highly anisotropic cryo-etching procedure for deep silicon structures, i.e. inductively coupled plasma reactive-ion etching (ICP-RIE)\cite{cryo1,cryo2,cryo3}. An example of a self-fabricated microstructure, manufactured with an Oxford\footnote{Oxford Instruments GmbH, Borsigstraße 15A, 65205 Wiesbaden, Germany} PlasmaPro 100 system at MNF Münster, is 
shown in fig. \ref{fig:etchresult}.
\footnotetext[11]{Münster Nanofabrication Facility, University of Münster, Busso-Peus-Str.10, 48149 Münster, Germany}
\footnotetext{JEOL (Germany) GmbH, Gute Änger 30, 85356 Freising, Germany}

Of course for a final design the hexagonal structure has to be etched through in order for the structure to become transparent. The sidewalls, which are currently 10\,$\mu$m thick, may need to be made thicker to provide enough stability for this transparent structure. However, transforming such a general honeycomb-like silicon structure into an MCP-like aTEF poses many more challenges \cite{Si-MCP2,Si-MCP}:  First, due to the high conductivity, the silicon structure needs to be covered by an insulating layer, e.g. SiO$_2$. Two additional layers need to be added to the side walls: First, a high-resistance layer to be able to apply the accelerating potential along the channel length, and then a secondary electron-emitting layer. The latter is not trivial, since the incoming electrons have a rather high energy of 18.6\,keV. Finally, a conductive layer must be applied to the front and back sides of the honeycomb structure to obtain a homogeneous potential distribution. If these challenges seem extremely high, it should still be noted that the aTEF should only allow a binary decision to distinguish between different electron cyclotron radii: The gain of the secondary cascade does not need to be nearly as homogeneous as in optical applications of MCPs and also the gain only needs to be in the range of 10 and not in the range $10^3-10^4$.

\begin{figure}[h]
 \centering 
 \includegraphics[width=\columnwidth]{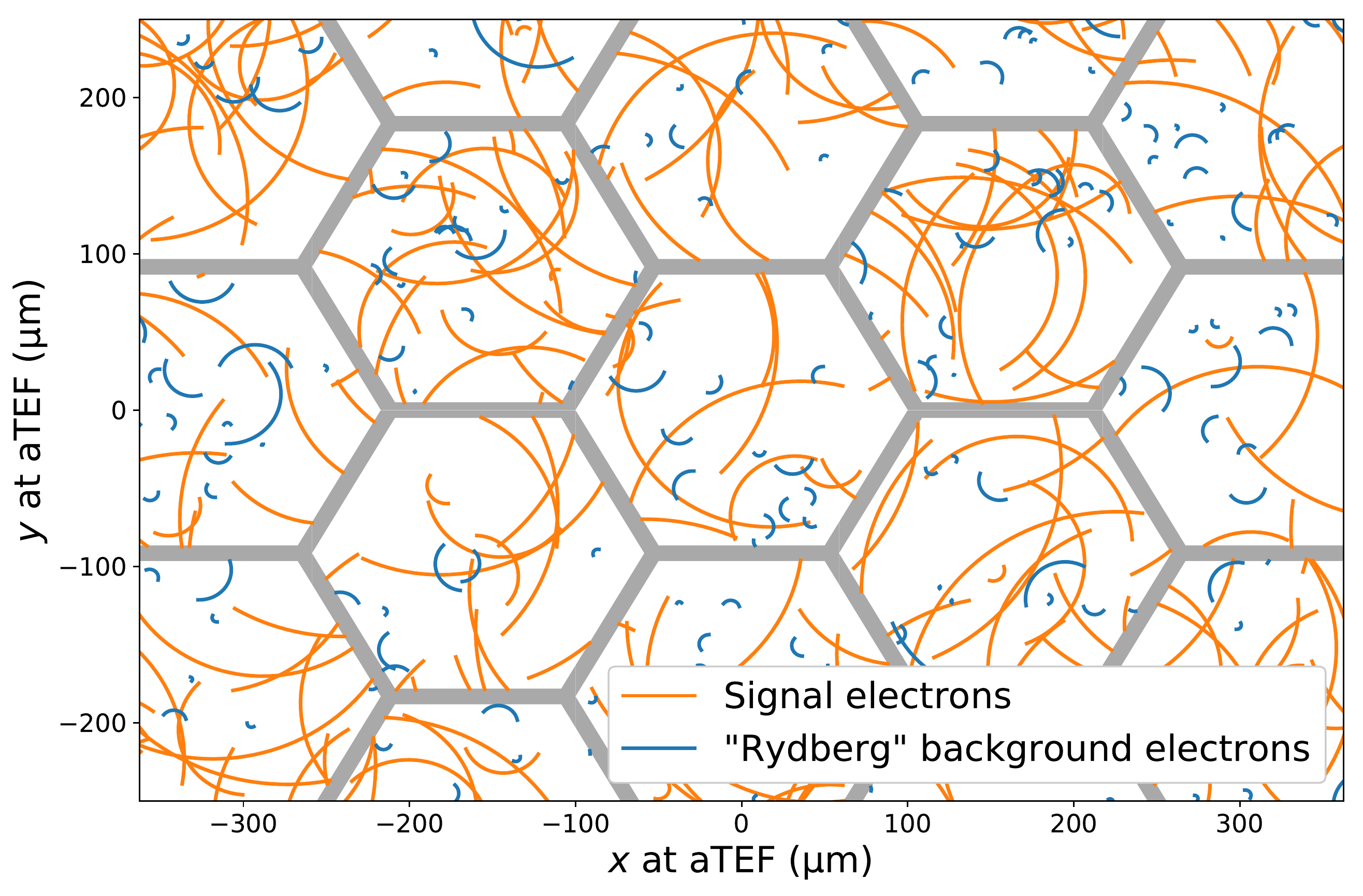}
 \caption{Simulation of signal (orange) and background (blue) electron tracks in a 2.4\,T axial magnetic field with the angular distributions shown in Fig. \ref{fig:background_at_KATRIN} for a honeycomb-like aTEF structure with an side length of $100\,\upmu$m, a wall thickness of $10\,\upmu$m, and a depth of $400\,\upmu$m. The tracking of an electron is stopped, when it either hits the aTEF surface or when it reaches the full depth without any hit.}
 \label{fig:aTEF_simulation} 
\end{figure}

A Monte Carlo simulation of electron tracks through the proposed geometry is illustrated in fig. \ref{fig:aTEF_simulation}.
%
%\SI{e5} 
Electrons are generated at the entrance of the three-dimensional hexagonal structure in a magnetic field of $2.4\,$T, with angles given by the angular distributions in fig.~\ref{fig:background_at_KATRIN} according to whether they are of ``Rydberg background'' type (blue tracks) or of ``signal'' type (orange tracks).
The magnetic field lines and, therefore, the guiding centers of the electrons' motion are directed perpendicular to the viewing plane. The tracking stops when an electron either hits the wall or reaches the full depth of the channel of \SI{400}{\micro\meter}.
Electrons are counted as ``detected'' in the former case and as ``not detected'' in the latter case. The percentage of detected electrons differs noticeably between the two types of electrons: $90\,\%$ of the signal electrons and $11.4\,\%$ of Rydberg background electrons interact with the wall. Taking into account an open-area ratio of 90\%, we therefore retain 81\% of the signal events and $10.3\,\%$ of the background events, corresponding to an improvement of the signal-to-background ratio by a factor of $\approx 7.9$.
\section{Conclusions}
We have proposed the concept of an aTEF to suppress electrons with low incident angles relative to the magnetic field axis and, thus, discriminate between different cyclotron radii.
Using a microchannel plate as a prototype for an active transverse energy filter, we successfully demonstrated this principle on a test setup with an angular-selective photoelectron source. The quantitative disagreement between the expected suppression factor and our test measurements can be explained by the angular width of our photoelectron source and the inappropriate geometry of the commercial MCP for our purpose. 

To obtain an estimate for the performance of an active transverse energy filter adapted to the needs of the KATRIN experiment, we performed Monte-Carlo simulations taking into account a hexagonal channel structure with an open area ratio of \SI{90}{\percent} and a channel diameter-to-length ratio of 1/4. With this geometry we can expect a significant improvement in the signal-to-noise ratio of nearly a factor 8 with an aTEF-type detector. 
To overcome the dark-count rates inherent to the glass substrates used in commercial MCPs, we investigated silicon as an intrinsically low-background material which would also be suitable for a semiconductor aTEF type detector. First tests to fabricate an appropriate microstructure in silicon have successfully been performed at the Münster Nanofabrication facility.
\section*{Acknowledgments}
We acknowledge the support of Ministry for Education and Research BMBF (contract number 05A20PMA) and Deutsche Forschungsgemeinschaft DFG (Research Training Group GRK 2149) in Germany. Support was also provided by the U.S. Department of Energy Office of Science, Office of Nuclear Physics under Award Number DE-FG02-97ER41020.
We further acknowledge the Münster Nanofabrication Facility (MNF) and especially Banafsheh Abasahl and Alexander Eich for their support during the fabrication of silicon microstructure samples. We kindly thank Norbert Wermes (University of Bonn) and Peter Lechner (Max Planck Semiconductor Laboratory, Munich) for their valuable inputs.
\section*{Declarations}
\paragraph{Funding}
This work was supported by the German Ministry for Education and Research BMBF (05A20PMA) and Deutsche Forschungs\-gemeinschaft DFG (Research Training Group GRK 2149).
\paragraph{Conflict of interest/Competing interests}
The authors have no relevant financial or non-financial interests to disclose.
\paragraph{Availability of data and materials}
The datasets generated during and/or analysed during the current study are available from the corresponding author on reasonable request.
\paragraph{Code availability}
Not available
\noindent
\bibliography{bibliography.bib}
\end{document}